\documentclass[modern]{aastex631}

\begin{document}

\title{Astrometric and photometric verification of faint blue white dwarfs 
in the \textit{Gaia} catalogue of nearby stars}

\correspondingauthor{Ralf-Dieter Scholz}
\email{rdscholz@aip.de}

\author[0000-0002-0894-9187]{Ralf-Dieter Scholz}
\affiliation{Leibniz Institute for Astrophysics Potsdam (AIP),\\
An der Sternwarte 16, 14482 Potsdam, Germany}

\begin{abstract}
	The \textit{Gaia} catalogue of nearby stars (GCNS) divided
	all objects with parallaxes $>$10\,mas into
	GCNS-selected and GCNS-rejected 100\,pc samples.
	Below the white dwarf (WD) sequence in the complete GCNS
	color-magnitude diagram (CMD), at $Gabs>14.7+4.7(G-RP)$,
	there appear 60 GCNS-selected faint blue white dwarfs (FBWDs).
	However this CMD region is also
	populated by 411 GCNS-rejected objects, mainly from 
	crowded regions towards the Galactic centre and the Magellanic 
	Clouds. 
	The WD catalog of \citet{2021MNRAS.508.3877G} lists only 47
        GCNS-selected but also 8 GCNS-rejected objects.
        I confirm 59 of the GCNS-selected but none of the GCNS-rejected
        objects as FBWDs from visual
        inspection and a proper motion check using additional optical sky
        surveys. Hence FBWDs form an additional branch in the CMD.
	Compared to the full GCNS-selected 100\,pc sample, FBWDs have
        relatively high proper motions and tangential velocities. They
        represent interesting targets for studies of ultracool or
        infrared-faint, and possibly also ultramassive WDs.
\end{abstract}


\section{Faint blue white dwarfs in the GCNS color-magnitude diagram} 

\textit{Gaia} color-magnitude diagrams (CMDs) reveal 
small numbers of faint blue white dwarfs (FBWDs) with
absolute \textit{Gaia} magnitudes of $Gabs \gtrapprox 15$\,mag
and a wide range of blue \textit{Gaia} colors.
These FBWDs may constitute
of ''ultracool'' \citep{2020MNRAS.499.1890M} 
or ''infrared-faint'' \citep{2020ApJ...898...84K}, 
but possibly also ''ultramassive" \citep{2021MNRAS.503.5397K} WDs.
The \textit{Gaia} catalogue of nearby stars 
\citep[GCNS;][]{2021A&A...649A...6G} is based on
the \textit{Gaia} Early Data Release 3 \citep[EDR3;][]{2021A&A...649A...1G}. 
Based on many \textit{Gaia} astrometric parameters and quality flags,
about 500000 EDR3 objects with measured magnitudes $Gmag$ and 
colors $G-RP$ and parallaxes $Plx>10$\,mas were divided into a
GCNS-selected 100\,pc sample of $\approx$296000 stars (59\%) and a 
GCNS-rejected sample of $\approx$204000 (41\%) objects.
Except for a few objects with extremely red $G-RP$ colors, all GCNS-selected
stars are shown in Fig.~\ref{fig:abcd}(a). In this CMD
the main sequence is well separated from the white dwarf (WD) sequence, 
which is mainly located within the marked color-magnitude box defined 
by \citet{2018MNRAS.480.3942H} for all nearby WDs within 20\,pc measured in 
the previous \textit{Gaia} DR2. However, 60 FBWDs appear
below this box, with absolute magnitudes
\begin{equation}
Gabs>14.7+4.7(G-RP).
\label{Eq1}
\end{equation}

The GCNS-rejected sample is
dominated by very faint objects towards the Galactic centre (GC) 
and the Magellanic 
clouds (MCs), where the \textit{Gaia} photometry, in particular in the $RP$ 
and $BP$ bands, and astrometry is strongly affected by image crowding.
Therefore, the tails of their $G-RP$ color distribution extend even outside
the blue and red limits shown in Fig.~\ref{fig:abcd}(b). Interestingly, there 
is an overlap of the blue tail of GCNS-rejected objects (411 objects 
following Eq.~\ref{Eq1}) with the above mentioned CMD region of FBWDs
that may throw discredit on the 60 GCNS-selected stars. On the other 
hand, only 55 WDs in the EDR3-based catalog of \citet{2021MNRAS.508.3877G} 
are found with Eq.~\ref{Eq1}, and 8 of them are members of the GCNS-rejected 
sample (Fig.~\ref{fig:abcd}(c)).

\section{Astrometric verification of FBWD candidates by proper motion check}

To investigate how much the apparent branch of FBWDs may be 
caused by \textit{Gaia} measuring errors, I inspected their
optical finder charts (including
own measurements in corresponding FITS images) and catalogs from 
different sources. The idea was to not only estimate the influence of crowding 
or close companions but confirm EDR3 proper motions above a chosen limit 
of 15\,mas/yr by combining the 
available \textit{Gaia} DR1-DR3 (epochs 2015-2016) positions with additional 
positional data. The presence of a confirmed proper motion was considered as 
supporting the EDR3 parallax, and consequently the absolute magnitude
of a FBWD candidate. However, I took
also into account the occurrence of other EDR3 sources at small
($\lessapprox$2\,arcsec) separations and EDR3 common proper motion (CPM) 
objects within 3\,arcmin and their available parallaxes. External data were 
taken from the APM \citep{1994IEEES...2...14I} and SuperCOSMOS 
\citep[SSS;][]{2001MNRAS.326.1279H} measurements of photographic Schmidt 
plates, the Sloan Digital Sky Survey \citep[SDSS;][]{2009ApJS..182..543A} and 
Pan-STARRS release 1 \citep[PS1;][]{2017yCat.2349....0C}. A proper motion
comparison was also made with the extended \textit{Gaia}-PS1-SDSS catalog 
\citep[GPS1+;][]{2020ApJS..248...28T}. In addition, I used the DESI Legacy 
Imaging Surveys \citep[][hereafter the Legacy Surveys]{2019AJ....157..168D}, 
but only in exceptional cases near-infrared data from the VISTA Hemisphere 
Survey \citep[VHS;][]{2013Msngr.154...35M}.

The EDR3 total proper motions $PM$ and numbers of visibility periods $Nper$
(an important astrometric quality parameters, in particular in
crowded regions)
of 411 GCNS-rejected objects are both systematically smaller than those 
of 60 GCNS-selected stars (Fig.~\ref{fig:abcd}(d)). Compared to the full 
sample of $\approx$302000 GCNS-selected stars with $Plx>10$\,mas, 
where 27\% have $PM$ around 
the peak of the distribution at 50$\pm$20\,mas/yr, only 15\% of the 60 
stars fall in this interval. 
Consequently, their tangential velocities are relatively
high (median $\approx$50\,km/s, maximum $\approx$160\,km/s). 

I confirmed 
the relatively high proper motions of all 60 GCNS-selected stars,
including the brightest (B = \object{Gaia EDR3 4049566372499936640})
and faintest (F = \object{Gaia EDR3 1674805012263764352}) of them
(Figs.~\ref{fig:abcd}(c,d)). With $Gabs \approx 17.3$\,mag, the latter
is nearly as faint as, but much bluer than, the unusual cool WD candidate 
\object{Gaia EDR3 6584418167391671808} \citep{2021RNAAS...5..229A}.
Only one of the 60 blue stars (H = \object{Gaia EDR3 1643819327188791040}) 
was previously classified as 
high-mass WD by \citet{2019ApJ...886..100C}. It has another WD
(\object{Gaia EDR3 1643819331484091136}) as CPM companion 
(separation 8.9\,arcsec) with similar parallax. 
The 60 objects are uniformly distributed over the 
sky, but one (W = \object{Gaia EDR3 4110333184616593280})
appears in PS1 and Legacy Surveys as a blue foreground object in the GC. 
It has a redder 
CPM companion separated by 4\,arcsec with a slightly smaller but more
precise parallax. Using the latter, the absolute magnitude
changes as indicated by an arrow in Fig.~\ref{fig:abcd}(c). Of 13 
GCNS-selected stars missing in \citet{2021MNRAS.508.3877G} I confirmed
12 as FBWDs. As seen in Fig.~\ref{fig:abcd}(d), half of them 
have $Nper < 12$ (one overlaps here with the unconfirmed object R, see below).

All eight WDs of \citet{2021MNRAS.508.3877G} in the GCNS-rejected sample
have very small $PM$ and $Nper$ (Fig.~\ref{fig:abcd}(d)) including five
objects in the GC and  MCs
regions and a quasar candidate (Q = \object{Gaia EDR3 6696431605961218176})
from \citet{2019MNRAS.490.5615B}. Only one
(C = \object{Gaia EDR3 5155424067737827968}) has a relatively high
$PM \approx 25$\,mas/yr that I confirmed using SDSS, PS1, and Legacy 
Surveys data. However, it has a large parallax error of about 3\,mas
and a CPM companion (separation 46\,arcsec) with a much smaller parallax
and ten times smaller error indicating a distance of 700\,pc. When
checking the proper motions of about 70 more GCNS-rejected objects with
$PM > 15$\,mas/yr, 85\% of which were in the crowded GC and MCs regions,
I did not find further FBWD candidates.

\section{Crowding and close companions affecting \textit{Gaia} color 
measurements}

\begin{figure*}[ht!]
\includegraphics[scale=0.75,angle=0]{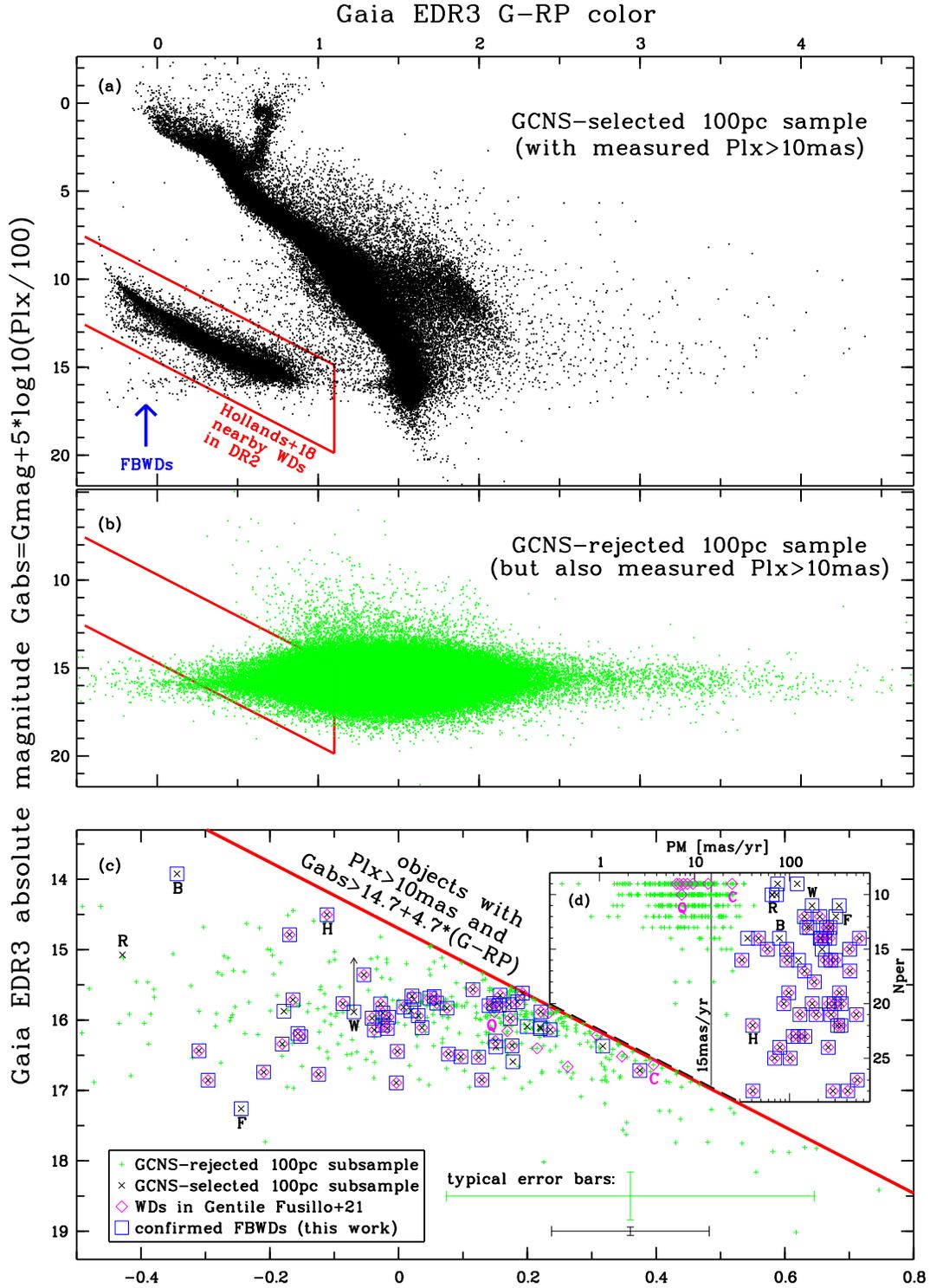}
	\caption{Color-magnitude diagram of full GCNS-selected \textbf{(a)} 
	and GCNS-rejected \textbf{(b)} 100\,pc samples. 
	The zoom in \textbf{(c)} 
	shows only the region of FBWDs (Eq.~\ref{Eq1}). The green and black
	error bars represent mean uncertainties of GCNS-rejected (green pluses)
	and GCNS-selected (black crosses) subsamples, respectively.
	Their total proper motions ($PM$) and numbers of visibility periods 
	($Nper$) are displayed in the inserted panel \textbf{(d)}. Objects 
	labelled in \textbf{(c)} and \textbf{(d)} are discussed in the text.
	\label{fig:abcd}}
\end{figure*}

The error bars shown in Fig.~\ref{fig:abcd}(c) are dominated by
$RP$ magnitude errors and relative parallax uncertainties. For
GCNS-rejected objects (green) they are larger than for GCNS-selected
stars (black). Most (53 of 60) GCNS-selected stars were also measured 
in PS1 or Legacy Surveys (22 in both). All but one of these appear blue 
on the corresponding finder charts. Surprisingly, the exception is
the object at the blue end of the shown CMD   
(R = \object{Gaia EDR3 6309477283343732096}), which in fact appears red 
in PS1 and VHS and is probably not a WD. It has a close (separation 3\,arcsec) 
brighter CPM companion 
($\Delta G \approx 5.8$\,mag but $\Delta J \approx 3.0$\,mag !)
with a similar parallax that obviously affected the $RP$ photometry.

\end{document}